\documentclass[prd,aps,twocolumn,nofootinbib,preprintnumbers,superscriptaddress,preprintnumbers,balancelastpage,longbibliography]{revtex4-2}
\usepackage[english]{babel}

\usepackage{amsmath,mathtools,physics,xfrac}
\usepackage[normalem]{ulem}
\usepackage{graphicx}
\usepackage{afterpage}
\usepackage{float}
\usepackage{rotating}
\usepackage{multirow}
\usepackage{tabularx}
\usepackage{booktabs}
\usepackage{fancyhdr}
\usepackage[utf8]{inputenc}
\usepackage{theorem}
\usepackage{moreverb}
\usepackage{euscript}
\usepackage{psfrag}
\usepackage{slashed}
\usepackage{mathtools}
\usepackage{makecell}
\usepackage{adjustbox}
\usepackage{dcolumn}
\usepackage{bm}
\usepackage[dvipsnames]{xcolor}
\usepackage{graphics}
\usepackage{hyperref}
\usepackage{lettrine}
\usepackage{amsfonts}
\usepackage{tikz}
\usetikzlibrary{decorations.text}
\input Zallman.fd

\LettrineTextFont{\itshape}
\definecolor{halpha}{HTML}{D100A0}

\hypersetup{
     colorlinks   = true,
     citecolor    = halpha,
     urlcolor     = halpha,
     linkcolor    = halpha
}

\usepackage{color,xcolor}

\definecolor{darkblue}{rgb}{0.0,0.0,0.75}
\definecolor{darkred}{rgb}{0.6,0.0,0}
\definecolor{darkgreen}{rgb}{0.0,0.6,0.}

\newcommand{\Ha}{\mathrm{H}\alpha}

\newcommand{\ega}{E_\gamma}

\newcommand\redsout{\bgroup\markoverwith{\textcolor{red}{\rule[0.5ex]{2pt}{0.4pt}}}\ULon}

\begin{document}
\usetikzlibrary{decorations.pathmorphing,calc}

\definecolor{gasblue}{RGB}{16,101,142}
\definecolor{waveblue}{RGB}{65,145,220}
\definecolor{palegas}{RGB}{216,232,247}
\definecolor{peach}{RGB}{244,163,112}
\definecolor{specblue}{RGB}{45,137,203}
\definecolor{magentaLine}{RGB}{205,77,191}

\preprint{SLAC-PUB-260710}

\title{Leading Axion Sensitivity with Interstellar Medium Optical Line Emission}

\author{Rebecca~K.~Leane}
\thanks{\href{mailto:rleane@stanford.edu}{rleane@stanford.edu}; \href{http://orcid.org/0000-0002-1287-8780}{0000-0002-1287-8780}
}
\affiliation{SLAC National Accelerator Laboratory, 2575 Sand Hill Rd, Menlo Park, CA 94025, USA}
\affiliation{Kavli Institute for Particle Astrophysics and Cosmology, Stanford University, Stanford, CA 94305, USA}

\author{Gowri Sundaresan}
\thanks{\href{mailto:gowrisun@stanford.edu}{gowrisun@stanford.edu}; \href{https://orcid.org/0000-0003-3417-7739}{0000-0003-3417-7739}
}
\affiliation{SLAC National Accelerator Laboratory, 2575 Sand Hill Rd, Menlo Park, CA 94025, USA}
\affiliation{Department of Physics, Stanford University, Stanford, CA, 94305, USA
}

\date{\today}

\begin{abstract}
We introduce interstellar optical line emission as a new probe of axion dark matter. Axion decays to photons or electrons can excite or ionize neutral hydrogen, producing H$\alpha$ emission at a fixed atomic wavelength rather than at an energy set by the axion mass. Developing the formalism and applying it to the Leo T galaxy with Very Large Telescope data, we derive strong constraints for eV--GeV scale axions, covering new regions of parameter space for both axion-photon and axion-electron couplings. These results establish optical line measurements as a new discovery channel for axion dark matter, probe poorly constrained parameter space below hydrogen's ionization threshold, and constrain axion explanations of JWST's high-redshift supermassive-black-hole observations.
\end{abstract}

\maketitle

\lettrine{D}{ark matter} (DM) is one of the clearest signs of physics beyond the Standard Model, and axions and axion-like particles are among its best-motivated candidates~\cite{Adams:2022pbo}. If axions are the DM, their cosmological survival implies extremely rare decays and tiny couplings. Astrophysical systems can nevertheless probe this regime, as they accumulate decays over enormous DM populations and long timescales. New observables exploiting this exposure are key to excluding or discovering axion DM. Conventional astrophysical searches for axion decay typically look for photons at energies set by the axion mass~\cite{Regis:2020fhw,Janish:2023kvi,Roy:2023omw,Pinetti:2025owq,Saha:2025any}. While powerful, such searches generally require dedicated analyses, or specialized reprocessing of astrophysical data, at each mass of interest. 

\begin{figure}
\vspace{-2mm}
    \centering
   \resizebox{1.04\linewidth}{!}{\begin{tikzpicture}[x=1cm,y=1cm,line cap=round,line join=round] 
  \path[use as bounding box] (0,0) rectangle (16.4,10);

  % Local figure colors
  \definecolor{gasblue}{RGB}{16,101,142}
  \definecolor{waveblue}{RGB}{65,145,220}
  \definecolor{palegas}{RGB}{216,232,247}
  \definecolor{peach}{RGB}{244,163,112}
  \definecolor{specblue}{RGB}{45,137,203}
  \definecolor{magentaLine}{RGB}{205,77,191}

  % ================================================================
  % (a) Axion-induced H-alpha emission from galaxy ISM
  % ================================================================
  \node[anchor=north west,align=center,
        font=\fontsize{16}{18.5}\selectfont\bfseries]
    at (0.55,9.5)
    {(a) Axion-induced \\ $\mathrm{H}\alpha$ emission};

% --- DM halo: single radial gradient (center -> edge), replacing glow loop + old shade
\shade[shading=radial, inner color=black, outer color=white, draw=none]
  (3.35,4.55) circle[radius=3.2];

% --- Curved "Axion DM Halo" label, placed to match the original top label location (~y=6.72)
% Uses an arc above the halo so the text sits in the same region as the old node at (3.35,6.72).
\path[
  decorate,
  decoration={
    text along path,
    text={|\color{black}\fontsize{15}{15}\selectfont|Axion DM Halo},
    text align={align=center},
    raise=0.9ex
  }
] (3.35,4.55) ++(150:2.18)
  arc[start angle=150, end angle=30, radius=2.18];

  % Neutral interstellar gas region
  \fill[palegas] (3.35,4.55) circle[radius=2];

  \node[gasblue,align=center,font=\fontsize{15}{15}\selectfont]
    at (3.35,5.30)
    {Neutral H gas};
    
\begin{scope}[yshift=0.2cm]
  % a -> gamma gamma
  \draw[gray!70,dashed,line width=1.0pt]
    (1.64,4.02) -- (2.09,4.02);
  \fill[gray!65] (2.09,4.02) circle[radius=0.052];

  \draw[waveblue,line width=1.4pt,
        decorate,decoration={snake,amplitude=0.052cm,segment length=0.16cm}]
    (2.13,4.03) -- (2.79,4.29);
  \draw[waveblue,line width=1.4pt,
        decorate,decoration={snake,amplitude=0.052cm,segment length=0.16cm}]
    (2.13,4.01) -- (2.79,3.76);

  \node[font=\fontsize{13}{18}\selectfont] at (1.82,3.72) {$a$};
  \node[font=\fontsize{13}{20}\selectfont] at (2.92,4.34) {$\gamma$};
  \node[font=\fontsize{13}{20}\selectfont] at (2.92,3.68) {$\gamma$};

  % a -> e+ e-
  \draw[gray!70,dashed,line width=1.0pt]
    (3.47,3.95) -- (3.91,3.95);
  \fill[gray!65] (3.91,3.95) circle[radius=0.052];
  \draw[waveblue,line width=1.25pt] (3.95,3.95) -- (4.49,4.27);
  \draw[waveblue,line width=1.25pt] (3.95,3.95) -- (4.49,3.70);

  \node[font=\fontsize{13}{13}\selectfont] at (3.67,3.67) {$a$};
  \node[anchor=west,font=\fontsize{13}{13}\selectfont]
    at (4.52,4.22) {$e^{+}$};
  \node[anchor=west,font=\fontsize{13}{13}\selectfont]
    at (4.52,3.64) {$e^{-}$};

\end{scope}

  % Escaping H-alpha photon
  \draw[red,line width=2.7pt,
        decorate,decoration={snake,amplitude=0.145cm,segment length=0.40cm}]
    (3.43,3.2) -- (3.43,0.72);

  \node[red,anchor=west,align=left,
        font=\fontsize{15}{15}\selectfont]
    at (3.72,1.52)
    {$\mathrm{H}\alpha\ (3\to2$\\[-1pt]
   transition)\\[-1pt]
   escapes};

  % ================================================================
  % (b) Search concept
  % ================================================================
 \begin{scope}[xshift=-0.2cm] 
   
  \node[anchor=north west,
        font=\fontsize{16}{18.5}\selectfont\bfseries]
    at (9.25,9.5) {(b) Search concept};

  \node[anchor=west,font=\fontsize{15}{14}\selectfont\bfseries]
    at (8,7.62) {Traditional decay-line searches:};
  \node[anchor=west,font=\fontsize{14}{13}\selectfont]
    at (8,7.10)
    {Line energy tracks $m_a$; limited reach};

  % Upper comparison axes
  \draw[->,line width=1.15pt] (8.30,5.02) -- (15.88,5.02);
  \draw[->,line width=1.15pt] (8.30,5.02) -- (8.30,6.55);
  \node[rotate=90,font=\fontsize{15}{18}\selectfont]
    at (7.96,5.78) {Flux};

  % Three mass-dependent line profiles
  \draw[peach,line width=1.55pt]
    (8.99,5.10)
      .. controls (9.16,5.13) and (9.29,5.57) .. (9.35,6.20)
      .. controls (9.42,5.58) and (9.55,5.15) .. (9.68,5.09);
  \draw[peach,line width=1.55pt] (9.35,5.06) -- (9.35,6.14);

  \draw[specblue,line width=1.55pt]
    (11.51,5.11)
      .. controls (11.70,5.14) and (11.82,5.59) .. (11.89,6.24)
      .. controls (11.96,5.60) and (12.08,5.16) .. (12.21,5.10);
  \draw[specblue,line width=1.55pt] (11.89,5.06) -- (11.89,6.18);

  \draw[magentaLine,line width=1.55pt]
    (14.04,5.10)
      .. controls (14.22,5.14) and (14.35,5.58) .. (14.42,6.20)
      .. controls (14.49,5.59) and (14.62,5.15) .. (14.76,5.09);
  \draw[magentaLine,line width=1.55pt] (14.42,5.06) -- (14.42,6.14);

  \node[font=\fontsize{12}{12}\selectfont] at (9.35,4.61) {$m_a(1)$};
  \node[font=\fontsize{12}{12}\selectfont] at (11.89,4.61) {$m_a(2)$};
  \node[font=\fontsize{12}{12}\selectfont] at (14.42,4.61) {$m_a(3)$};

  \node[gray!60,font=\fontsize{14}{14}\selectfont]
    at (12.28,4.02) {vs.};

  \node[anchor=west,font=\fontsize{15}{15}\selectfont\bfseries]
    at (8,3.49) {ISM $\mathrm{H}\alpha$ line search:};
  \node[anchor=west,font=\fontsize{14}{14}\selectfont]
    at (8,2.98)
    {Fixed wavelength; broad $m_a$ reach};

  % Lower fixed-wavelength search axes
  \draw[->,line width=1.15pt] (8.30,0.90) -- (15.88,0.90);
  \draw[->,line width=1.15pt] (8.30,0.90) -- (8.30,2.43);
  \node[rotate=90,font=\fontsize{15}{18}\selectfont]
    at (7.96,1.66) {Flux};

  \draw[red,line width=1.65pt]
    (11.03,0.98)
      .. controls (11.19,1.01) and (11.32,1.46) .. (11.39,2.12)
      .. controls (11.46,1.47) and (11.59,1.02) .. (11.72,0.97);
  \draw[red,line width=1.65pt] (11.39,0.94) -- (11.39,2.06);

  \node[red,font=\fontsize{12}{12}\selectfont]
    at (11.39,0.43) {$\mathrm{H}\alpha\ (656\,\mathrm{nm})$};
\end{scope}
    
\end{tikzpicture}}
   \vspace{-6mm}
    \caption{
    Schematic of axion-induced $\Ha$ emission and the search strategy. Axion decay products can ionize or excite neutral H in the interstellar medium, producing detectable $\Ha$ line emission. A single $\Ha$ measurement can therefore simultaneously probe a wide range of axion masses $m_a$.
    }
    \label{fig:Schematic}
    \vspace{-4mm}
\end{figure}
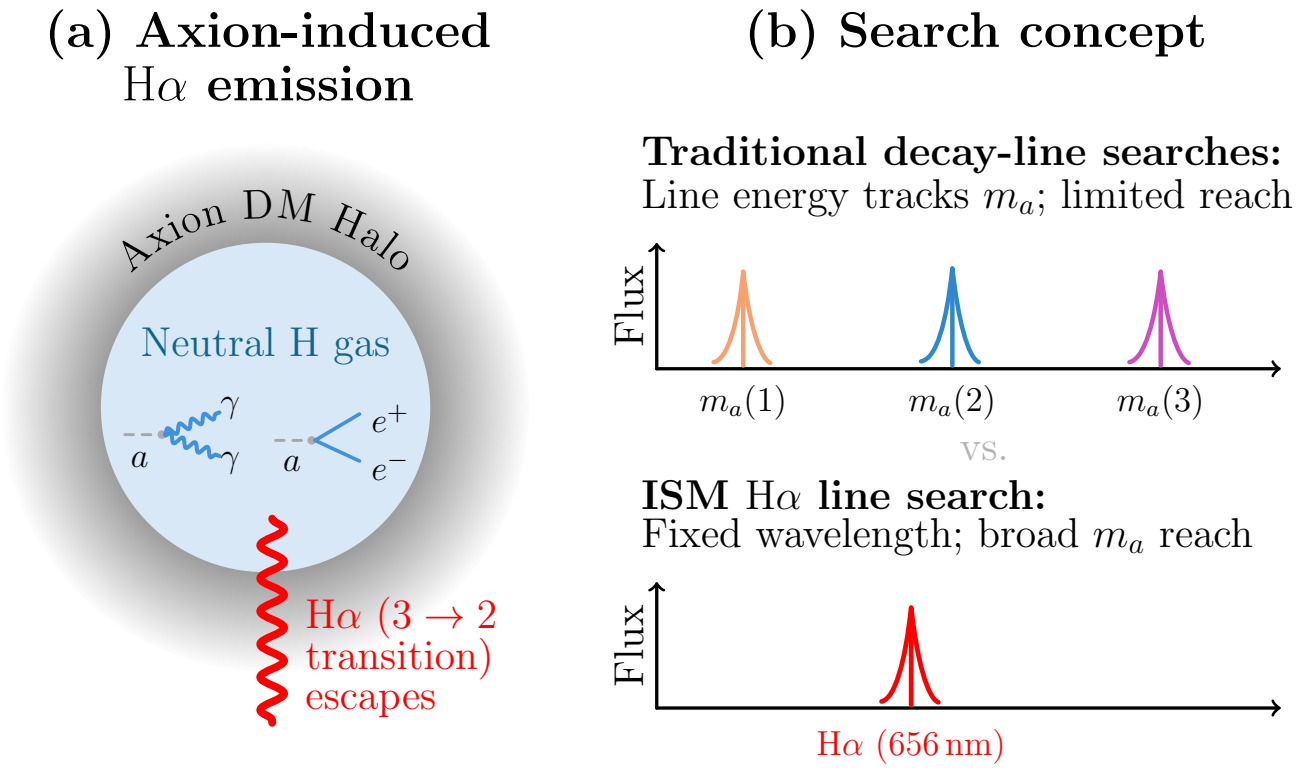

In this \textit{Letter}, we introduce a new signature for axions. If axion decay products excite or ionize neutral hydrogen (H) in the interstellar medium (ISM), the energy is reprocessed into de-excitation radiation at fixed atomic wavelengths, largely independent of the axion mass. A single line measurement therefore probes a wide range of axion masses, as illustrated in Fig.~\ref{fig:Schematic}. Gas heating~\cite{Chivukula:1989cc,Bhoonah:2018wmw,Wadekar:2019mpc,Wadekar:2021qae,Wadekar:2022ymq} and molecular cloud ionization~\cite{Prabhu:2022dtm,Blanco:2023bgz,DelaTorreLuque:2025zjt,Perez:2026cqi} also use the ISM as a DM detector, but infer DM energy deposition indirectly from gas temperature or chemistry; we instead directly detect radiation from DM-induced atomic transitions. We use $\Ha$, the $n=3\rightarrow2$ transition of H ($E_{\Ha}=1.89$ eV), as it (i) is efficiently produced in radiative de-excitation, (ii) escapes neutral gas without Ly$\alpha$-like resonant trapping and with little dust extinction, and (iii) is a standard tracer of ionization in galaxies, so deep observations already exist across many systems. Gas-rich dwarf galaxies, with large DM densities and clean environments, are ideal targets; we focus on Leo T, a nearby, DM-dominated dwarf with substantial neutral gas and negligible star formation.

Ref.~\cite{Leane:2025ohu} previously proposed $\Ha$ as a probe of DM, considering H ionization from generic DM energy injection. Here we develop the formalism for axions, tracking photon and electron decay products, and include for the first time DM-induced $\Ha$ from excitation. Excitation enhances reach overall, as it produces $\Ha$ directly, whereas ionization does so only after the resulting H$^+$ recombines with an electron; at the smallest couplings, where recombination has not equilibrated, excitation dominates. Importantly, it also opens sensitivity below H's ionization threshold. This is precisely where axions~\cite{FriedlanderSchonVincent2023,Lu:2024zwa,Aggarwal:2025pit,Lu:2025kbe,Xu:2024vdn} could explain JWST's surprising high-redshift supermassive black holes~\cite{Larson2023CEERS,Harikane2023BLAGN,Fan2023CosmicDawn}, whose formation remains a mystery~\cite{BrommLoeb2003FirstSMBH,ShangBryanHaiman2010}. Applying our formalism to Leo T data from the Multi Unit Spectroscopic Explorer (MUSE) on the Very Large Telescope (VLT), where extended $\Ha$ emission was not observed~\cite{Vaz_2023}, we derive leading constraints on axion couplings to photons and electrons over complementary mass ranges. We also exclude part of the axion parameter space that can explain JWST's early black holes, and show that future $\Ha$ observations can probe it further.\\

\noindent\textbf{\textit{H$\bm\alpha$ Emission from Excited and Ionized Gas.---}}
Energy deposited in neutral H can produce $\Ha$ emission through two channels: ionization and excitation. We now discuss each in the context of generic energy injection.

\textit{Ionization.}---Above the H ionization threshold, an incident photon $\gamma$ or electron $e$ can ionize H. The resulting H$^+$ can then recombine with a free electron into an excited state, which subsequently de-excites. When Lyman photons from the de-excitation cascades are trapped and locally reprocessed (called ``Case-B recombination''), as expected for the optically thick Leo~T ISM, $f_{\Ha}\simeq0.46$ of recombinations produce an $\Ha$ photon~\cite{RydenPogge2021_IISM}.

\textit{Excitation.}---Incident particles can also directly excite H from its $1s$ ground state to an upper state $n\ell$. Photon absorption peaks at Lyman-series energies, extending to nearby energies via the finite line width, while electrons can excite H to any level with excitation energy below their kinetic energy. We define $P_{\Ha}(i)$ as the probability that the radiative cascade from an excited state $i$ contains a $3\rightarrow2$ ($\Ha$) transition, and obtain it recursively as
\begin{equation}
P_{\Ha}(i)=
\sum_{f<i}B_{if}
\left[
\delta_{n_i,3}\delta_{n_f,2}
+
\left(1-\delta_{n_i,3}\delta_{n_f,2}\right)
P_{\Ha}(f)
\right],
\label{eq:PHalpha-recursion}
\end{equation}
with $P_{\Ha}(i)=0$ for $n_i\leq2$. The first term counts direct $3\rightarrow2$ decays; the second follows the cascade from daughter state $f$. The effective branching fraction is
\begin{equation}
B_{if}=\frac{A_{i\rightarrow f}\,\bigl(1-\delta_{f,1s}\bigr)}
{\sum_{f'<i}A_{i\rightarrow f'}\,\bigl(1-\delta_{f',1s}\bigr)}\,,
\label{eq:radiative-branching-fraction}
\end{equation}
where $A_{i\rightarrow f}$ are the Einstein coefficients for electric-dipole-allowed decays~\cite{NIST_ASD}. The factor $(1-\delta_{f,1s})$ removes Lyman decays $np\rightarrow1s$; the Leo~T gas is optically thick to these photons, which lie below the H and He ionization thresholds, so they are resonantly reabsorbed rather than escaping, ionizing, or heating the gas. Each $np$ excitation is thus recycled until it decays through a non-Lyman channel, analogous to Case-B recombination.

The initially excited populations depend on the projectile $k\in\{\gamma,e\}$ and its energy $E_k$ through the state-resolved excitation cross sections $\sigma^k_{1s\rightarrow n\ell}(E_k)$. Photoexcitation obeys electric-dipole selection rules and so populates only $np$ states ($\ell=1$), with
\begin{equation}
\sigma^\gamma_{1s\rightarrow n\ell}(E_\gamma)
=
\frac{2\pi^2\alpha}{m_e}\,
f_{1s\rightarrow n\ell}\,
\phi_{1s\rightarrow n\ell}(E_\gamma)\,,
\label{eq:photoexcitation-cross-section}
\end{equation}
where $\alpha$ is the fine structure constant, $m_e$ is the electron mass, $f_{1s\rightarrow n\ell}$ is the oscillator strength~\cite{NIST_ASD}, and $\phi_{1s\rightarrow n\ell}$ is the normalized
line profile. Electron impact can populate a broader set of $\ell$ states, and we take $\sigma^e_{1s\rightarrow n\ell}(E_e)$ from Ref.~\cite{CCCDatabase}. The probability of producing an $\Ha$ photon, averaged over initially populated states, is then
\begin{equation}
P_{\Ha}^{k}(E_k)
=
\frac{
\sum_{n\ell}
\sigma^k_{1s\rightarrow n\ell}(E_k)\,
P_{\Ha}(n\ell)
}{
\sum_{n\ell}
\sigma^k_{1s\rightarrow n\ell}(E_k)
}\,,
\
k\in\{\gamma,e\}\,.
\label{eq:PHalpha-projectile}
\end{equation}
Together, $f_{\Ha}$ for ionization and $P^k_{\Ha}(E_k)$ for excitation determine the $\Ha$ yield per deposited energy. Further details of the line profiles, atomic-data inputs, and cascade examples are in the Supplemental Material.\\

\noindent\textbf{\textit{Axion-Induced H$\bm\alpha$ Flux.---}} In environments where DM co-exists with neutral gas, such as in Leo T, axion decays can source the photons and electrons described above, leading to $\Ha$ production. For an observing aperture $A$, the $\Ha$ flux from decays of axion DM of mass $m_a$, with partial width $\Gamma_{a\to kk}$ to final state $k\in\{\gamma,e\}$, is
\begin{equation}
\mathcal{F}_{\Ha} [A] = \frac{ E_{k}^{\rm dec}\,\Gamma_{a\to kk}}{4\pi \,m_a} \left(f^{\Ha}_{{\rm exc},k}+f^{\Ha}_{{\rm ion},k} \,f_{{\rm eq},k}\right)\mathcal{D}[A]\,,
\label{eq:Flux}
\end{equation}
where the total kinetic energy in the two decay products is $E_{\gamma}^{\rm dec}=2E_{\gamma}=m_a$ for photon-decay and $E_{e}^{\rm dec}=2\,E_e=m_a-2\,m_e$ for electron-positron decay.  The factors $f^{\Ha}_{{\rm exc},k}$ and $f^{\Ha}_{{\rm ion},k}$ are the fractions of this energy converted into $\Ha$ photons via excitation and ionization of H; these are derived in the next two sections. Since ionization yields $\Ha$ only once the resulting H$^+$ recombines, the ionization term carries a factor $f_{{\rm eq},k}\leq1$, the fraction of the equilibrium recombination rate reached over the quiescent time. For the bulk of the parameter space $f_{{\rm eq},k}\simeq1$, and it falls below unity only at the lowest ionization rates (see Supplemental Material), where excitation then dominates. Finally, $\mathcal{D}[A]$ is the D-factor,
\begin{equation}
\mathcal{D}[A]\equiv \int_A d\Omega\!\int_{\rm los} dl\,\rho_a\,,
\label{eq:D factor}
\end{equation}
where  $d\Omega$ is the differential solid angle, $l$ is the line-of-sight distance, and $\rho_a$ is the axion DM density. 

The partial width for axion decay to an electron-positron pair, for axion-electron coupling $g_{aee}$, is~\cite{Bauer_2017}
\begin{equation}
\Gamma_{a \to e^- e^+} \;=\; \frac{g_{aee}^2\, m_a}{8\pi}\,
\sqrt{1 - \frac{4m_e^2}{m_a^2}}\;\Theta(m_a - 2m_e)\,.
\label{eq:ALP to e+e-}
\end{equation}
Radiative corrections from tree-level $g_{a\gamma\gamma}$ are negligible here over the MeV--GeV range~\cite{Langhoff_2022}. For decay to photons via axion-photon coupling $g_{a\gamma\gamma}$, the partial width including radiative corrections from the electron coupling is~\cite{Bauer_2017}
\begin{align}
    \Gamma_{a\to\gamma\gamma}&=\frac{m_a^{3}}{64\pi}\,\left|\,g_{a\gamma\gamma}
-\frac{\alpha g_{aee}}{\pi \,m_e}\left[1-\tau f^{2}(\tau)\right]\right|^{2} \,,\label{eq:ga2gammafull}\\
   f(\tau)&=
    \begin{cases}
    \arcsin\!\left(\dfrac{1}{\sqrt{\tau}}\right), & (\tau\ge 1)\\
    \left(\dfrac{\pi}{2}+\dfrac{i}{2}\ln\dfrac{1+\sqrt{1-\tau}}{1-\sqrt{1-\tau}}\right), & (\tau<1)
    \end{cases} \label{eq: ftau}
\end{align}
where $\tau=(2m_e/m_a)^2$. For $m_a \ll m_e$, the loop term is suppressed by $(m_a/m_e)^2$, so for nonzero tree-level $g_{a\gamma\gamma}$,
\begin{equation}
\Gamma_{a\to\gamma\gamma} \;=\; \frac{m_a^3\, g_{a\gamma\gamma}^2}{64\pi}\, .
\label{eq:A to 2 gamma}
\end{equation}
We also consider photophilic (electrophobic) axions, with vanishing tree-level $g_{aee}$, and photophobic (electrophilic) axions, with suppressed tree-level $g_{a\gamma\gamma}$ (for motivation for such models, see $e.g.$ Refs.~\cite{Di_Luzio_2020,Sokolov_2021,Eberhart:2025lyu, Craig_2018, Farina_2017, Darm__2021, Langhoff_2022}).  In the photophobic case, the electron coupling radiatively induces an effective photon coupling, given by Eq.~(\ref{eq:ga2gammafull}) with $g_{a\gamma\gamma}=0$. In the photophilic case, the photon coupling radiatively induces an effective electron coupling. Assuming no fine-tuned cancellations, this is, up to an $\mathcal{O}(1)$ factor~\cite{Dessert_2019},
\begin{equation}
   |g_{aee,\rm eff}| \simeq 5\times10^{-4}\,\frac{2\pi m_e}{\alpha}\,|g_{a\gamma\gamma}|\,.
   \label{eq:Photophilic gaee-eff}
\end{equation}

\noindent\textbf{\textit{Axion Energy Deposition in H$\bm\alpha$ via Decay to Electron-Positron Pairs.---}}
For axions decaying to $e^+e^-$, the $\Ha$ yield depends on how the injected electrons and positrons (hereafter electrons) propagate and lose energy in the neutral gas. We consider an observing aperture $A$ within the warm neutral medium (WNM) core of the Leo~T galaxy. Because the electron propagation length can exceed the aperture size, electrons and their secondaries produced inside $A$ may escape before depositing their energy, while those produced outside can diffuse in. We account for both effects with the two-zone transport model of Ref.~\cite{Leane:2025ohu}, which gives the fraction $F_A(E_e)$ of electron power injected throughout the WNM core that is deposited collisionally within the aperture; accordingly, for $e^\pm$ decay $\mathcal{D}[A]$ in Eq.~\eqref{eq:Flux} is replaced by the D-factor of the full WNM core (see Supp. Material). The ionization and excitation contributions are then
\begin{align}
f^{\Ha}_{{\rm ion},e}(E_e) &= F_A(E_e)\,\frac{f^{\rm H}_{\rm ion}(E_e)}{I_{\rm H}}\,f_{\Ha}\,E_{\Ha}\,,
\label{eq:fdepe_mono}\\
f^{\Ha}_{{\rm exc},e}(E_e) &= F_A(E_e)\,\frac{f^{\rm H}_{\rm exc}(E_e)}{E^e_{\rm exc}(E_e)}\,P^e_{\Ha}(E_e)\,E_{\Ha}\,,
\label{eq:fdep_electro-excitation}
\end{align}
where $f^{\rm H}_{\rm ion}$ and $f^{\rm H}_{\rm exc}$ are the fractions of electron kinetic energy deposited into H ionization and excitation in a mixed H/He gas~\cite{Furlanetto_2010}, $I_{\rm H}=13.6$~eV is the H ionization energy, and $P^e_{\Ha}$ is given by Eq.~(\ref{eq:PHalpha-projectile}). Each ratio converts deposited energy into a number of ionizations or excitations, which is then weighted by the probability of producing an $\Ha$ photon. The mean excitation energy is
\begin{equation}
E^{e}_{\rm exc}(E_e)=\frac{\sum_{n\ell}\sigma^{e}_{1s\to n\ell}(E_e)\,\Delta E_{1s\to n\ell}}{\sum_{n\ell}\sigma^{e}_{1s\to n\ell}(E_e)}\,,
\label{eq:Eexc_electroexcitation}
\end{equation}
with $\Delta E_{1s\to n\ell}$ the energy of the $1s\to n\ell$ transition.\\

\noindent\textbf{\textit{Axion Energy Deposition in H$\bm\alpha$ via Decay to Photons.---}}
The probability that an axion-decay photon is absorbed within the aperture is
\begin{equation}
P_{\rm abs}(E_\gamma)=1-e^{-\tau(E_\gamma)}\, ,
\label{eq: Pabs of photons}
\end{equation}
where $\tau$ is the optical depth,
\begin{equation}
\tau(E_\gamma)
= \rho_{\rm gas}\, \ell_{\rm gas}
\sum_{i\in\{\mathrm{H},\mathrm{He}\}}
w_i\left(\frac{\mu_{\rm en}}{\rho}\right)_{i}(E_\gamma)\,,
\label{eq: photon opacity}
\end{equation} 
where $\rho_{\rm gas} \approx 2.4 \times 10^{-25} \rm g/\rm cm^{3}$ is the average in-aperture gas density, $\ell_{\rm gas}\simeq 101\ \mathrm{pc}$ is the characteristic photon path length through the Leo~T gas in the MUSE aperture~\cite{Leane:2025ohu}, $w_{\rm H}=0.75$ and $w_{\rm He}=0.25$ are the H and He mass fractions, and $(\mu_{\rm en}/\rho)_i$ are the mass energy-absorption coefficients. For $10.2\leq E_\gamma<13.6$~eV, photons can only excite H, so $(\mu_{\rm en}/\rho)_{\rm H}=\sum_{n\ell}\sigma^\gamma_{1s\to n\ell}(E_\gamma)/m_{\rm H}$, with $m_{\rm H}$ the H mass, and $(\mu_{\rm en}/\rho)_{\rm He}$ is negligible. At higher energies we use Ref.~\cite{1990A&A...237..267B} for $13.6\leq E_\gamma<30$~eV, Ref.~\cite{CXRO:atten2} for $30\leq E_\gamma<1000$~eV, and Ref.~\cite{HubbellSeltzer_XAAMDI_2004} above 1~keV (see Supp. Material).

Above threshold, the ionization contribution is~\cite{Leane:2025ohu}
\begin{align}\label{eq:fdep_gamma_Ha}
f^{\Ha}_{{\rm ion},\gamma}(E_\gamma) &= P_{\rm abs}(E_\gamma)\,f_{\Ha}\,\frac{E_{\Ha}}{E_\gamma}\,\Theta(E_\gamma-I_{\rm H})\\
&\times\Bigg[1+\sum_{I_i<E_\gamma}\! f_i(E_\gamma)\,(E_\gamma-I_i)\,\frac{f^{\rm H}_{\rm ion}(E_\gamma-I_i)}{I_{\rm H}}\Bigg],\nonumber
\end{align}
where the sum runs over species with ionization energy below $E_\gamma$ ($I_{\rm H}=13.6$~eV, $I_{\rm He}=24.6$~eV), and
\begin{equation}
f_i(E_\gamma)=
\frac{w_i\,(\mu_{\rm en}/\rho)_i(E_\gamma)} 
{\sum_{j\in\{\mathrm{H},\mathrm{He}\}} w_j\,(\mu_{\rm en}/\rho)_j(E_\gamma)},
 i\in\{\mathrm{H},\mathrm{He}\}
\end{equation}
is the fraction of photon absorptions occurring on species $i$. The leading term counts one H ionization per absorbed photon: either direct H photoionization, or He photoionization followed by He recombination photons below the He threshold that ionize H~\cite{RydenPogge2021_IISM}. The sum counts secondary H ionizations by the resulting photoelectrons, with $f_{\rm ion}^{\rm H}$ denoting the fraction of their kinetic energy deposited into H ionization.

\begin{figure*}[t]
\centering
\begin{minipage}[t]{0.48\linewidth}
  \vspace{0pt}\includegraphics[width=\linewidth]{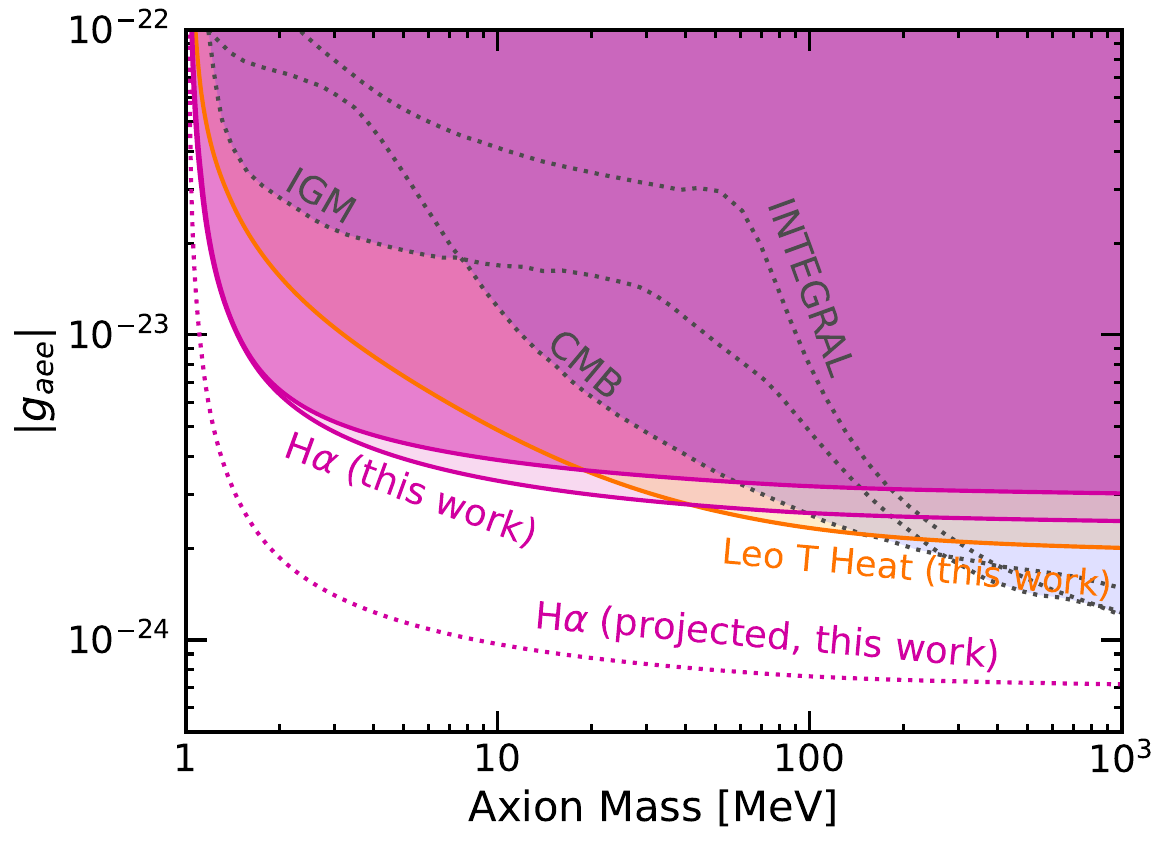}
\end{minipage} \hspace{1em}
\begin{minipage}[t]{0.48\linewidth}
  \vspace{0pt}\includegraphics[width=\linewidth]{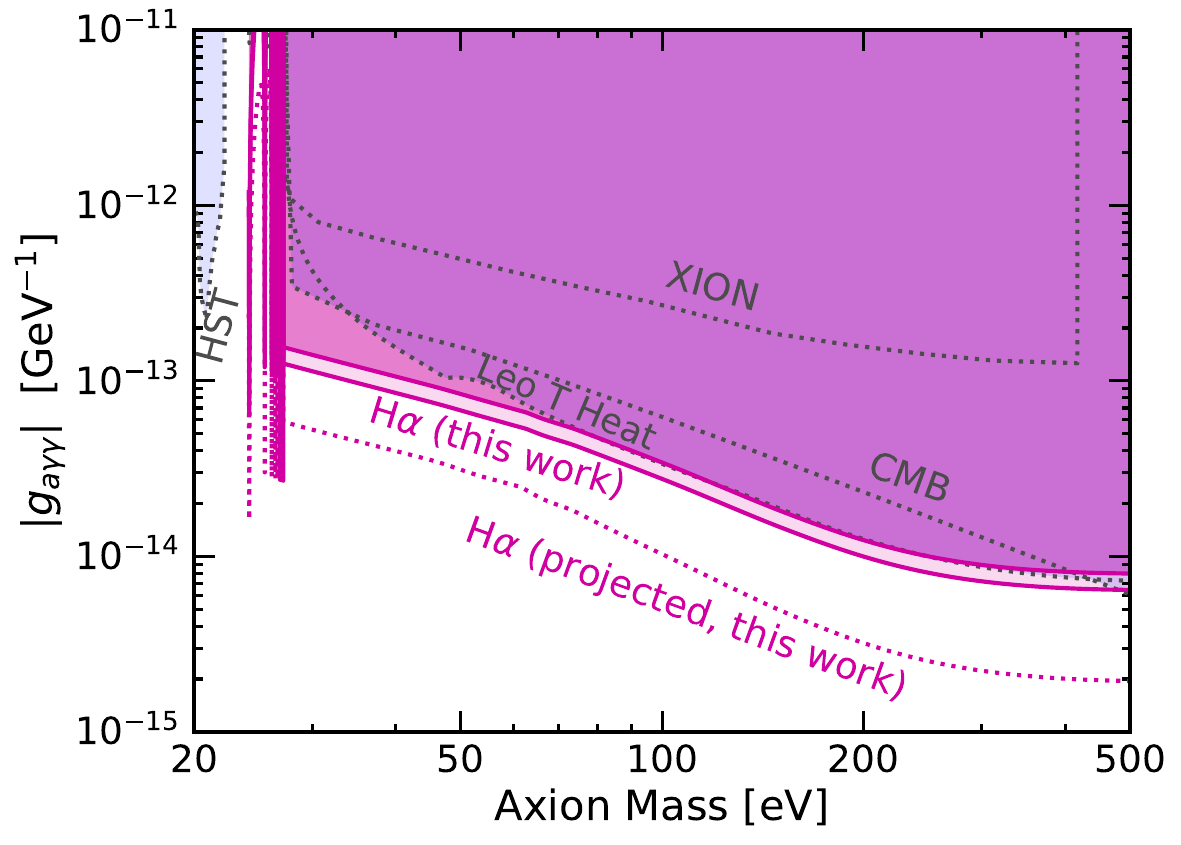}
\end{minipage}
\vspace{-6pt}
\caption{95\% C.L. limits on axion-electron coupling $|g_{aee}|$ from $a\to e^+e^-$ (left) and axion-photon coupling $|g_{a\gamma\gamma}|$ from $a\to\gamma\gamma$ (right) from MUSE/VLT $\Ha$ observations of Leo~T (solid; lighter/darker shading for NFW/Burkert profiles) and ELT/HARMONI projections (dotted). Orange: our Leo~T heating limit; gray: existing limits (see text). The narrow features at $24.2\lesssim m_a<27.2$~eV (right) are Lyman-series photoexcitation resonances, shown in detail in Fig.~\ref{fig:SMBH ROI}.}
    \label{fig:axion couplings}
\end{figure*}

Similarly, the excitation contribution is
\begin{align}\label{eq: photo-excitation full}
&f^{\Ha}_{{\rm exc},\gamma}(E_\gamma) = P_{\rm abs}(E_\gamma)\,\frac{E_{\Ha}}{E_\gamma}\Bigg[P^\gamma_{\Ha}(E_\gamma)\,\Theta(I_{\rm H}-E_\gamma)\\
&+\sum_{I_i<E_\gamma}\! f_i(E_\gamma)\,(E_\gamma-I_i)\,\frac{P^e_{\Ha}(E_\gamma-I_i)\,f^{\rm H}_{\rm exc}(E_\gamma-I_i)}{E^e_{\rm exc}(E_\gamma-I_i)}\Bigg],\nonumber
\end{align}
where the first term is direct photoexcitation of H, and the second is collisional excitation by photoelectrons.\\

\noindent\textbf{\textit{New Constraints on Axions and Future Sensitivities.---}} 
We translate the predicted axion-induced $\Ha$ flux into axion mass and coupling limits using the non-observation of extended $\Ha$ emission from Leo~T with MUSE/VLT~\cite{Vaz_2023}. We require the flux in Eq.~\eqref{eq:Flux} not to exceed the aperture-integrated 95\% C.L. limit, $\mathcal{F}_{\Ha}^{\rm lim}=7.8\times10^{-18}~{\rm erg\,s^{-1}\,cm^{-2}}$~\cite{Vaz_2023,Leane:2025ohu}. Dust extinction and additional $\Ha$ attenuation are negligible for Leo~T~\cite{deJong:2008qi,Leane:2025ohu}. We also calculate illustrative ELT/HARMONI projections, assuming ten Leo~T-like dwarfs and a factor-of-five improvement in flux sensitivity relative to MUSE, motivated by the expected $\sim25\times$ gain in observing speed~\cite{HARMONI_Performance_Oxford}. These projections are representative; a dedicated forecast should account for HARMONI's smaller field of view and target-to-target variation in gas and DM distributions.

Figure~\ref{fig:axion couplings} (left) shows our limits on the axion-electron coupling from $a\to e^-e^+$.  Our $\Ha$ constraint is strongest in the $\mathcal{O}(1$--$10)$ MeV range, where it improves existing bounds by nearly an order of magnitude; the projected ELT/HARMONI results extend the reach to $|g_{aee}|\sim10^{-24}$.  We also show new Leo~T heating limits obtained using the methodology of Ref.~\cite{Leane:2025ohu} (see also Ref.~\cite{Wadekar:2021qae}), and applied to this axion parameter space for the first time. The intergalactic medium (IGM), cosmic-microwave background (CMB), and INTEGRAL limits are recast from DM lifetime bounds~\cite{Liu_2021,Slatyer:2016qyl,Cirelli:2023tnx} via Eq.~\eqref{eq:ALP to e+e-}. Further limits from Voyager~1 $e^\pm$ data and the Galactic 511~keV line exist~\cite{Boudaud_2017,De_la_Torre_Luque_2025,DelaTorreLuque:2023cef,Nguyen:2025tkl}, but these depend on uncertain cosmic-ray and positron propagation.

\begin{figure}[t]
    \includegraphics[width=\linewidth
    ]{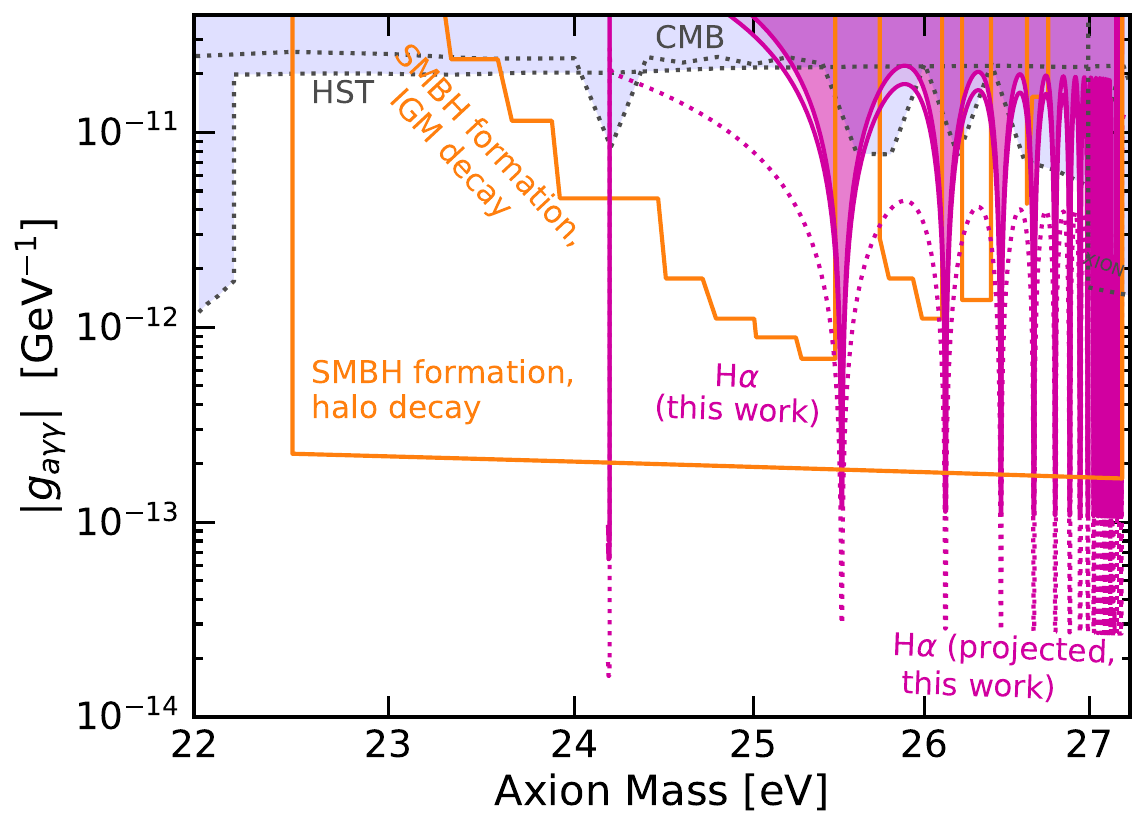}
    \vspace{-10pt}
\caption{95\% C.L. MUSE/VLT $\Ha$ limits (solid; lighter/darker shading for NFW/Burkert profiles) and ELT/HARMONI projections (dotted) on $|g_{a\gamma\gamma}|$, compared with axion parameter space proposed for direct-collapse SMBH formation via halo~\cite{Lu:2024zwa} and IGM~\cite{Aggarwal:2025pit} decays (above orange lines). Existing limits are shown~\cite{Todarello:2024qci,Xu:2024vdn,Carenza_2023,Nakayama_2022,O_Hare_GitHub}.}
    \label{fig:SMBH ROI}
\end{figure}

Figure~\ref{fig:axion couplings} (right) shows our limits on $|g_{a\gamma\gamma}|$ from $a\to\gamma\gamma$. The sharp features at low masses (shown in detail in Fig.~\ref{fig:SMBH ROI}) occur where axion-decay photons are resonantly absorbed by ground-state H on Lyman-series lines, $1s\to np$ with $m_a=2E_{1s\to np}$; the excited atom then cascades down, emitting $\Ha$ ($3\to2$) along the way. On these resonances, accessible only through excitation, our limits lie up to two orders of magnitude below HST dwarf-galaxy bounds~\cite{Todarello:2024qci}. Constraints from gamma-ray attenuation~\cite{Bernal_2023} and extragalactic background light~\cite{Arias:2012az} are not shown, as they do not account for absorption by neutral hydrogen above $m_a \simeq 20.4$~eV (see Ref.~\cite{Bernal_2023}). Above the ionization threshold, our limits also improve on bounds from extra early-universe ionization (XION)~\cite{Cadamuro_2012}, CMB anisotropies (recast from the lifetime limits of Ref.~\cite{Xu:2024vdn} via Eq.~\eqref{eq:A to 2 gamma}), and Leo~T heating~\cite{Wadekar:2021qae,Leane:2025ohu}. The projected ELT/HARMONI results further extend the sensitivity.

\begin{figure}[t]
    \centering
    \includegraphics[width=\linewidth]{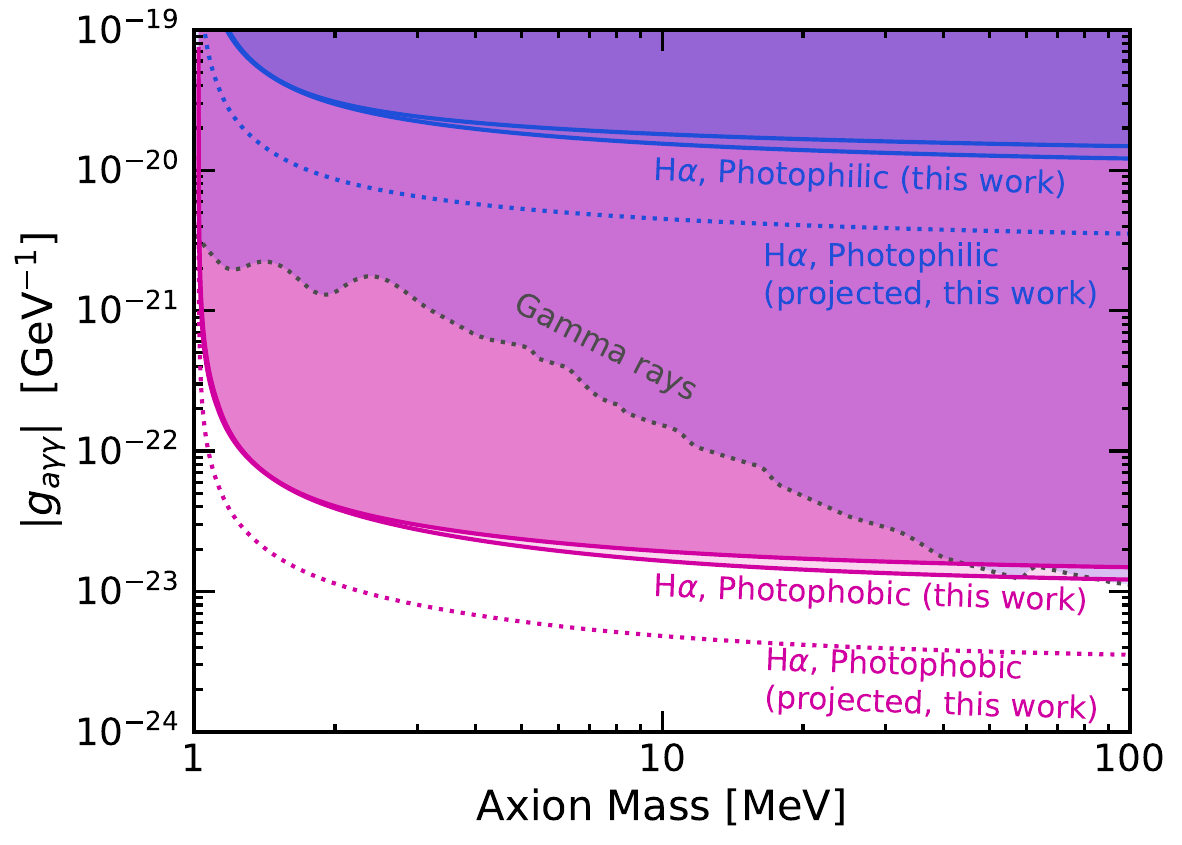}
    \vspace{-10pt}
\caption{95\% C.L. limits on $|g_{a\gamma\gamma}|$ for photophilic (blue) and photophobic (pink) axions from MUSE/VLT $\Ha$ observations of Leo~T (solid; lighter/darker shading for NFW/Burkert profiles) and ELT/HARMONI projections (dotted). For photophobic axions, $g_{a\gamma\gamma}$ is the effective coupling induced by $g_{aee}$. Gamma-ray limits (shaded) apply to both models.} 
    \label{fig:ALP-ga2gamma-eff}
\end{figure}

Figure~\ref{fig:ALP-ga2gamma-eff} recasts our $\Ha$ limits on $a\to e^+e^-$ for two axion models. For photophilic (electrophobic) axions, $g_{a\gamma\gamma}$ radiatively induces an effective $g_{aee}$ (Eq.~\eqref{eq:Photophilic gaee-eff}); for photophobic (electrophilic) axions, $g_{aee}$ instead induces an effective $g_{a\gamma\gamma}$ (Eq.~\eqref{eq:ga2gammafull}). INTEGRAL and COMPTEL searches for $a\to\gamma\gamma$ (``gamma rays'')~\cite{Calore:2022pks,Essig:2013goa} constrain the diphoton rate directly and so apply to both models (shaded); they remain stronger than $\Ha$ in the photophilic case. In the photophobic case, $\Ha$ improves on gamma-ray limits by up to two orders of magnitude. Since Eq.~\eqref{eq:ga2gammafull} rescales all $a\to e^+e^-$ limits by the same factor at each mass, the model-dependent comparison with the other electron-coupling limits in Fig.~\ref{fig:axion couplings} (left) carries over directly, and $\Ha$ provides the leading bound for $m_a\sim1$–$50$~MeV.\\

\noindent\textbf{\textit{Implications for Axion Explanations of JWST's High-Redshift Supermassive Black Holes.---}} 
JWST has uncovered a large population of supermassive black holes (SMBH) at high redshift~\cite{Larson2023CEERS,Harikane2023BLAGN}, intensifying the open question of how massive black-hole seeds formed so early in cosmic history~\cite{Fan2023CosmicDawn}. One proposed mechanism is direct collapse, in which Lyman--Werner (LW) photons with energies $11.2$--$13.6$~eV dissociate molecular H, suppressing gas cooling and fragmentation and enabling massive seeds to form~\cite{BrommLoeb2003FirstSMBH,ShangBryanHaiman2010}. Decaying axion DM has recently been proposed as a source of this radiation~\cite{FriedlanderSchonVincent2023,Lu:2024zwa,Aggarwal:2025pit,Lu:2025kbe,Xu:2024vdn}. We point out that the same decays that dissociate molecular H in the early Universe would also excite neutral H in nearby gas-rich dwarfs today, producing testable $\Ha$ emission.

Figure~\ref{fig:SMBH ROI} compares our $\Ha$ limits with the axion parameter space proposed for LW-assisted direct collapse, from decays in high-redshift halos~\cite{Lu:2024zwa} and in the intergalactic medium~\cite{Aggarwal:2025pit}. For halo decays, we show the case of negligible H$_2$ self-shielding at $z\sim20$, which Ref.~\cite{Lu:2024zwa} expects to be most realistic and which extends to the smallest couplings; stronger shielding requires couplings a few times larger, which are more strongly constrained by our work.  For intergalactic decays, we show the $z=10$ benchmark of Ref.~\cite{Aggarwal:2025pit}. Our limits exclude parts of the parameter space for both scenarios; axion explanations of JWST's early black holes are therefore already testable through local $\Ha$ emission, and ELT/HARMONI-like observations can probe them further.

In Fig.~\ref{fig:SMBH ROI}, our sensitivity is greatest on the Lyman-series lines, where axion-decay photons are resonantly absorbed by ground-state H. Slightly off resonance, a decay photon instead scatters through a virtual excited state into a real lower state, and the photon emitted in this scattering is shifted in energy by the decay photon's offset from resonance (Raman scattering). For Ly$\beta$, the only $\Ha$-producing decay is $3p\to2s$, so $\Ha$ itself is shifted; since the MUSE data analysis counts only $\Ha$ within $\pm7.7$~\AA\ of the line~\cite{Vaz_2023}, our sensitivity is therefore confined near line center. For higher lines, every $\Ha$-producing path begins with an infrared photon that carries the shift, so $\Ha$ remains in band. As the shifted photon is itself a narrow line, a line search across the full optical spectrum rather than at $\Ha$ alone could extend the sharp Ly$\beta$ sensitivity across its wings, further probing the SMBH parameter space. We model the lines with Voigt profiles; a full Kramers--Heisenberg treatment could modify the limits away from line centers (see Supplemental Material).\\

\noindent\textbf{\textit{Summary and Outlook.---}} We introduced interstellar optical line emission as a new, leading probe of axion DM. Axion decays to photons or $e^+e^-$ can ionize or excite neutral H, producing $\Ha$ emission. Unlike conventional decay searches, which look for photons at an energy set by the axion mass, this signal appears at a fixed atomic transition. A single optical-line measurement can thus test broad regions of axion parameter space.

We developed the formalism for axion-induced $\Ha$ emission from both photon and electron decays, including excitation and ionization channels, and applied it to the non-observation of extended $\Ha$ emission from Leo~T with MUSE/VLT, yielding leading sensitivity to axion-photon and axion-electron couplings in complementary mass ranges. For $a\to\gamma\gamma$, resonant excitation opens the difficult-to-probe region below the H ionization threshold, where we improve existing limits on $|g_{a\gamma\gamma}|$ by up to two orders of magnitude, while remaining competitive up to $m_a\sim1$~keV. For $a\to e^+e^-$, our $\Ha$ and new Leo~T heating constraints provide leading sensitivity across the $\sim1$--$100$~MeV range, improving existing bounds by nearly an order of magnitude, and for photophobic axions, $\Ha$ improves on direct gamma-ray limits by up to two orders of magnitude. Sensitivity to sub-ionization-threshold photons also allows us to exclude part of the axion parameter space proposed to generate the Lyman--Werner radiation required for direct-collapse explanations of JWST's high-redshift black holes, and future $\Ha$ observations offer a way to test it further.

The reach of this approach can be extended through deeper Leo~T observations, larger samples of quiet gas-rich dwarfs, and ELT-class spectroscopy. Other H transitions, including higher Balmer and infrared lines, could provide cross-checks and extend accessible parameter space. A signal would be distinguished by its line ratios and by morphology tracing the DM distribution, without the metal lines of conventional ionized gas. Existing and future optical spectroscopy therefore provide a new route to powerful advances in axion DM sensitivity.
\\

\noindent\textbf{\textit{Acknowledgments.---}}
We thank Ian Holst, Giuseppe Lucente, and Linda Xu for helpful discussions, and especially Tracy Slatyer for helpful discussions and suggesting the inclusion of DM-induced excitation. RKL and GS are supported by the U.S. Department of Energy under Contract DE-AC02-76SF00515.

\clearpage
\newpage
\onecolumngrid
\begin{center}
\textbf{\large Leading Axion Sensitivity with Interstellar Medium Optical Line Emission}

\vspace{0.05in}
{ \it \large Supplemental Material}\\ 
\vspace{0.05in}
{Rebecca K. Leane and Gowri Sundaresan}
\end{center}
\onecolumngrid
\setcounter{equation}{0}
\setcounter{figure}{0}
\setcounter{section}{0}
\setcounter{table}{0}
\setcounter{page}{1}
\makeatletter
\renewcommand{\theequation}{S\arabic{equation}}
\renewcommand{\thefigure}{S\arabic{figure}}
\renewcommand{\thetable}{S\arabic{table}}

\tableofcontents

\section{Axion Lagrangian}
In this work, we consider an axion that constitutes all of the DM, with a Lagrangian of the form
\begin{equation}
\mathcal{L} \supset \frac{1}{2}\left(\partial_\mu a\right)^2
-\frac{1}{2} m_a^2 a^2
-\frac{1}{4}g_{a\gamma\gamma}\,a\,F_{\mu\nu}\tilde F^{\mu\nu}
+\frac{g_{aee}}{2m_e}(\partial_\mu a)\,\bar e\,\gamma^\mu\gamma_5\,e \, ,
\label{eq:axion Lagrangian}
\end{equation}
where $m_a$ is the axion mass, $F_{\mu\nu}$ is the electromagnetic field-strength tensor, and $\tilde F^{\mu\nu}\equiv \frac{1}{2}\epsilon^{\mu\nu\rho\sigma}F_{\rho\sigma}$ is its dual. The couplings $g_{a\gamma\gamma}$ and $g_{aee}$ parameterize the axion interactions with photons and electrons, respectively; $e$ is the electron field, and $m_e$ is the electron mass. We remain agnostic about the ultraviolet origin of the axion mass and couplings, and use ``axion'' generically to denote any pseudoscalar DM candidate described by Eq.~\eqref{eq:axion Lagrangian}, without distinguishing between the QCD axion and more general axion-like particles.

\section{Leo T Environment and MUSE/VLT Observations}
\label{sec: Leo T}

Leo T is a faint, gas-rich, DM-dominated dwarf galaxy in the Local Group. It lies beyond the Milky Way virial radius, approximately $420~\mathrm{kpc}$ from the Galactic center. Leo T has been extensively studied through 21-cm observations and deep optical spectroscopy \cite{Faerman:2013pmm,Patra_2018,Ryan-Weber:2007guk,Adams_2018,Vaz_2023}. In particular, Giant Metrewave Radio Telescope (GMRT) and Westerbork Synthesis Radio Telescope (WSRT) observations have mapped its gas distribution and kinematics \cite{Ryan-Weber:2007guk,Adams_2018}, while Hubble Space Telescope (HST) and Multi Unit Spectroscopic Explorer (MUSE) observations have constrained its stellar population and star-formation history~\cite{Weisz_2012,Vaz_2023}.

The neutral gas in the core of Leo T contains both a cold neutral medium (CNM) and a warm neutral medium (WNM). The CNM accounts for approximately $10\%$ of the neutral-H mass, whereas the WNM dominates the volume \cite{Adams_2018}. MUSE spectroscopy finds no extended $\Ha$ emission down to very low surface brightness, consistent with the extremely low present-day star-formation activity \cite{Vaz_2023}. We conservatively restrict our analysis to the WNM, whose properties are better constrained.

The MUSE Wide Field Mode has a $1'\times1'$ field of view. At a distance of approximately $420~\mathrm{kpc}$, this corresponds to a square aperture with half-side $a=61~\mathrm{pc}$, centered on Leo T. We model the WNM core as a sphere of radius $R=350~\mathrm{pc}$, corresponding to a central line-of-sight depth of
$L_{\rm depth}\simeq 2R=700~\mathrm{pc}$.

A small ionized fraction is expected even within the predominantly neutral WNM. Ionization in the outer regions of Leo T is driven primarily by the metagalactic UV background, while soft X-rays can penetrate into the inner regions~\cite{Wadekar:2021qae,Wadekar:2019mpc}. Although the metagalactic UV flux is uncertain, the X-ray background relevant to the inner WNM is comparatively well constrained, limiting the corresponding uncertainty in the central electron density~\cite{Haardt_2012}.

The ionization fraction enters our calculation in two ways. First, when evaluating the equilibrium factor $f_{{\rm eq},k}$ in Eq.~(\ref{eq:Flux}), we conservatively neglect the ambient electron population in the equilibration timescale. Including these electrons would shorten the equilibration time and strengthen our results. Second, we evaluate the electron energy-deposition fractions $f_{\rm ion}^{\rm H}(E_e)$ and $f_{\rm exc}^{\rm H}(E_e)$ using the results of Ref.~\cite{Furlanetto_2010} for an ionization fraction $x_e=0.01$. These quantities enter Eqs.~(\ref{eq:fdepe_mono}), (\ref{eq:fdep_electro-excitation}), (\ref{eq:fdep_gamma_Ha}), (\ref{eq: photo-excitation full}). A smaller ionization fraction would increase the energy deposited into H ionization and excitation, and hence increase the resulting $\Ha$ yield, making our choice conservative.

The DM distribution in Leo T is constrained by H\,{\sc i} kinematics and dynamical modeling of its gas and stars \cite{Ryan-Weber:2007guk,Adams_2018,Faerman:2013pmm,Patra_2018}. These observations robustly constrain the enclosed mass over the radii traced by the data, but do not uniquely determine the profile at arbitrary radii. The principal uncertainty is whether the inner profile is cored or cusped. We therefore consider two standard spherical parameterizations
\begin{equation}
    \text{Burkert:}\qquad
\rho_a(r)=\frac{\rho_0\,r_0^{3}}{(r+r_0)\,(r^{2}+r_0^{2})} ~,
\qquad
\text{NFW:}\qquad
\rho_a(r)=\frac{\rho_s}{(r/r_s)\,\bigl(1+r/r_s\bigr)^2} ~,
\label{eq: Leo T profile}
\end{equation}
with $\rho_0 = 5.4 \rm ~GeV/\rm cm^{3}$ and $r_0 = 400 \rm pc$ for the Burkert profile, and $\rho_s = 2.9 \rm ~GeV/\rm cm^{3}$ and $r_s = 350 \rm pc$ for the NFW profile, following Ref.~\cite{Leane:2025ohu}. These profiles are used to evaluate the decay D-factor in Eq.~(\ref{eq:D factor}).

\begin{figure}[t]
    \includegraphics[width=0.4\linewidth]{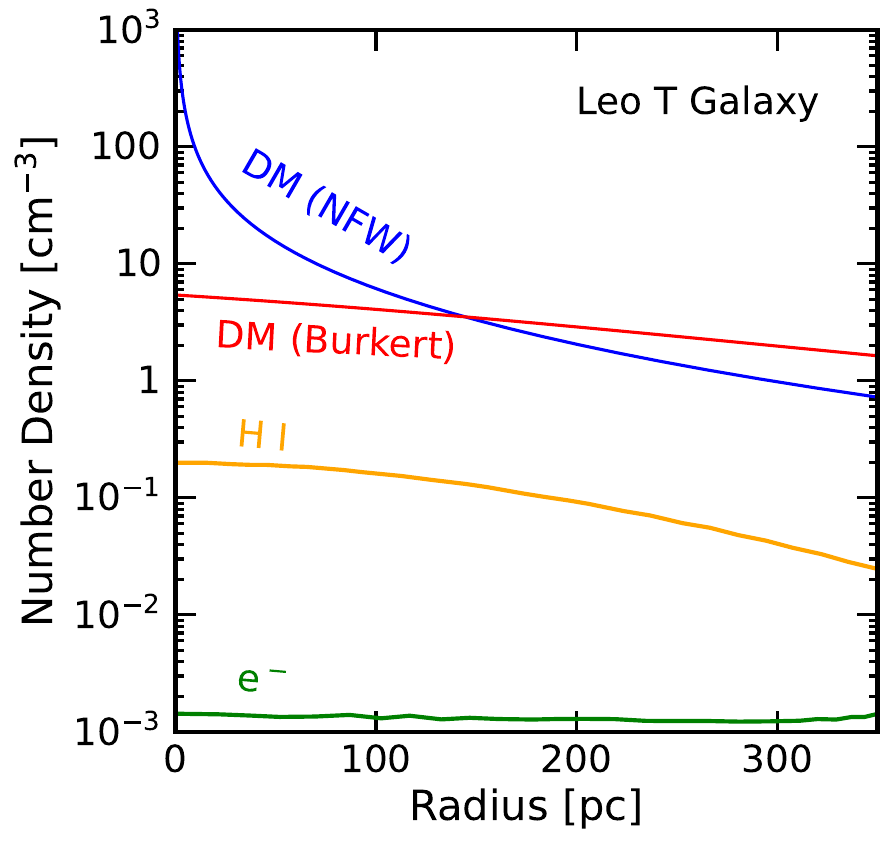}
    \caption{Number densities in the Leo T dwarf: atomic H, electrons, and DM (assuming 1 GeV mass) with NFW and Burkert profiles are shown. This figure is reproduced from Ref.~\cite{Leane:2025ohu}.}
    \label{fig:Leo T profile}
\end{figure}

Figure~\ref{fig:Leo T profile} shows the corresponding DM and gas-density profiles of Leo T.

\section{Radiative Cascade and $P_{\Ha}(i)$ Calculation Examples}

Here we give examples of the cascade probability $P_{\Ha}(i)$ defined in Eq.~(\ref{eq:PHalpha-recursion}), the probability that the radiative cascade from an initially excited state $i$ contains a $3\rightarrow2$ transition. The initial state may be populated by either photo-excitation or electron-impact excitation. Throughout, Lyman decays $np\rightarrow1s$ are removed from the branching fractions, since the Leo~T gas resonantly reabsorbs these photons, so each $np$ excitation is recycled until it decays through a non-Lyman channel. Branching between decay channels is set by the Einstein coefficients $A_{i\rightarrow f}$ for spontaneous emission~\cite{NIST_ASD}, as in Eq.~(\ref{eq:radiative-branching-fraction}).

States with $n\leq2$ cannot produce $\Ha$, so $P_{\Ha}(2s)=P_{\Ha}(2p)=0$. The $3s$ and $3d$ states can only decay to $2p$ (electric-dipole selection rules require $\Delta\ell=\pm1$), so
\begin{equation}
P_{\Ha}(3s)=P_{\Ha}(3d)=1\,.
\end{equation}
The $3p$ state can decay to $1s$ or $2s$. With the Lyman channel removed, only $3p\rightarrow2s$ remains, giving
\begin{equation}
P_{\Ha}(3p)=1\,.
\label{eq:PHalpha-3p-LeoT}
\end{equation}

Similarly, $4p$ decays to $2s$, $3s$, or $3d$ (the $4p\rightarrow3p$ decay is forbidden, $\Delta\ell=0$), giving
\begin{equation}
P_{\Ha}(4p)=\frac{A_{4p\rightarrow3s}\,P_{\Ha}(3s)+A_{4p\rightarrow3d}\,P_{\Ha}(3d)}{A_{4p\rightarrow2s}+A_{4p\rightarrow3s}+A_{4p\rightarrow3d}}\,.
\label{eq:PHalpha-4p}
\end{equation}
Higher states build on these results. For example,
\begin{equation}
P_{\Ha}(5p)=\frac{A_{5p\rightarrow3s}P_{\Ha}(3s)+A_{5p\rightarrow3d}P_{\Ha}(3d)+A_{5p\rightarrow4s}P_{\Ha}(4s)+A_{5p\rightarrow4d}P_{\Ha}(4d)}{A_{5p\rightarrow2s}+A_{5p\rightarrow3s}+A_{5p\rightarrow3d}+A_{5p\rightarrow4s}+A_{5p\rightarrow4d}}\,,
\label{eq:PHalpha-5p}
\end{equation}
which requires $P_{\Ha}(4s)$ and $P_{\Ha}(4d)$ as inputs. These examples generalize to the recursion relation in Eq.~(\ref{eq:PHalpha-recursion}).

\section{Axion $\Ha$ Energy Deposition via Electron-Positron Decay: Supporting Material}
\label{sec: electron dep}

\subsection{Two-Zone Transport Model for Energy Deposition Within the Observing Aperture}

Here we summarize the two-zone electron-transport model developed in Ref.~\cite{Leane:2025ohu}. The model determines the fraction $F_A(E_e)$ of axion-injected electron power ultimately deposited through collisions within the observing aperture $A$. This quantity enters Eqs.~(\ref{eq:fdepe_mono}) and (\ref{eq:fdep_electro-excitation}) of the main text.

For $a\to e^-e^+$, the $\Ha$ yield is governed by competition between collisional energy losses, radiative losses, and diffusive transport. Although the observing aperture is smaller than the WNM core, electrons produced outside the aperture can diffuse into it and deposit energy. We therefore divide the WNM into a central aperture zone $A$ and the surrounding region $B$. The exterior of the WNM core, where we conservatively neglect DM injection, is denoted zone $C$.

Electrons diffuse through magnetic turbulence characterized by the Alfven speed $v_A$ and coherence length $L_{\rm coh}$, and a diffusion coefficient $K$~\cite{1966ApJ...146..480J,2002cra..book.....S,2022SSRv..218...33E},
\begin{align}
v_A &= \frac{B_{\rm mag}}{\sqrt{4\pi\,\rho_{\rm ion}}}\,,\qquad
K = \frac{1}{3}\,v_A\,L_{\rm coh}\,,
\label{eq:diffusion}
\end{align}
where $B_{\rm mag}$ is the magnetic-field strength and $\rho_{\rm ion}$ is the ion mass density. For Leo T, we use the ion density shown in Fig.~\ref{fig:Leo T profile}, a uniform field $B_{\rm mag}=1~\mu{\rm G}$, and $L_{\rm coh}=50~{\rm pc}$. Coherence lengths in dwarf galaxies may be as small as approximately $1$--$20~{\rm pc}$ \cite{1996ApJ...458..194M}, making our choice conservative.

The diffusive escape rate from zone $X$ to a neighboring zone $Y$ is approximated by \cite{Redner_2023}
\begin{align}
\Gamma_{X\to Y}^{\rm esc}
&=
\frac{\pi^2K}{\ell_{X\rightarrow Y}^2},
\end{align}
where $\ell_{X\rightarrow Y}$ is the effective distance to the relevant boundary. For the MUSE/VLT observations of Leo T, we adopt
\begin{equation}
\ell_{A\rightarrow B}
=
\frac{a}{\sqrt{2}}
\simeq 43~{\rm pc},
\
\ell_{B\rightarrow A}
=
60~{\rm pc}\ \text{(NFW)},
\
70~{\rm pc}\ \text{(Burkert)},
\
\ell_{B\rightarrow C}
=
125~{\rm pc}\ \text{(NFW)},
\
100~{\rm pc}\ \text{(Burkert)}.
\end{equation}
The first value corresponds to a conservative choice of the escape rate within our approximate two-zone transport model. Direct escape from $A$ to $C$ is negligible and is therefore omitted. For electrons produced in $B$, the effective distances are DM-weighted harmonic means of the local distances to the corresponding boundaries.

In zone $X$, the collisional-deposition and radiative-loss rates are
\begin{align}
\Gamma_X^{\rm dep}(E_e)
&= \frac{c}{E_e}\sum_{i\in\{\mathrm{H},\mathrm{He}\}} m_i\,n_{i,X}\,S_i^{\rm coll}(E_e),
\\
\Gamma_X^{\rm rad}(E_e)
&= \frac{c}{E_e}\sum_{i\in\{\mathrm{H},\mathrm{He}\}} m_i\,n_{i,X}\,S_i^{\rm rad}(E_e)
+ \Gamma_X^{\rm IC+syn}(E_e).\nonumber
\end{align}
Here $S_i^{\rm coll}$ and $S_i^{\rm rad}$ are the corresponding stopping powers \cite{NIST_SRD124_Landing}, and $m_i$ and $n_{i,X}$ are the mass and number density of species $i$. The term $\Gamma_X^{\rm IC+syn}$ accounts for inverse-Compton and synchrotron losses, which are negligible in the parameter space considered. The total loss rate is $\Gamma_X^{\rm tot}=\Gamma_X^{\rm dep}+\Gamma_X^{\rm rad}$; only the collisional component contributes to the $\Ha$ signal.

In zone $B$, the inverse mean residence time is
\begin{equation}
\Lambda_B(E_e)=\Gamma_B^{\rm tot}(E_e)+\Gamma_{B\to A} ^{\rm esc}+\Gamma_{B\to C} ^{\rm esc}\,,
\end{equation}
and the probability of reaching $A$ before losing energy or escaping into $C$ is
\begin{equation}
r_B(E_e)=\frac{\Gamma_{B\to A} ^{\rm esc}}{\Lambda_B(E_e)}\,.
\end{equation}
For an electron initially in $A$, the probability that its energy is ultimately deposited collisionally within $A$ is
\begin{equation}
P_{A\to A}(E_e)=
\frac{\Gamma_A^{\rm dep}(E_e)}{\Gamma_A^{\rm tot}(E_e)+\Gamma_{A\to B} ^{\rm esc}\big[1-r_B(E_e)\big]}\,.
\end{equation}
The second term in the denominator accounts for electrons that leave $A$ and do not return; direct escape from $A$ to $C$ is negligible.

The total fraction of injected electron power deposited collisionally within the aperture is therefore
\begin{equation}
F_A(E_e)=
f_A\,P_{A\to A}(E_e)+
(1-f_A)\,r_B(E_e)\,P_{A\to A}(E_e)\,,
\label{eq:FA_main}
\end{equation}
where $f_A$ is the fraction of axion-injected $e^\pm$ produced in $A$, with the remaining fraction $1-f_A$ produced in $B$, compared to the whole WNM core. As in Ref.~\cite{Leane:2025ohu}, $F_A$ is therefore combined with the D-factor of the full WNM core, rather than $\mathcal{D}[A]$, in Eq.~(\ref{eq:Flux}) for $e^\pm$ decay.

\subsection{Electron-Impact Excitation Calculation: Treatment of Levels and Validation of Energy Deposition}

\begin{figure}[h]
    \centering
    \includegraphics[width=0.5\linewidth]{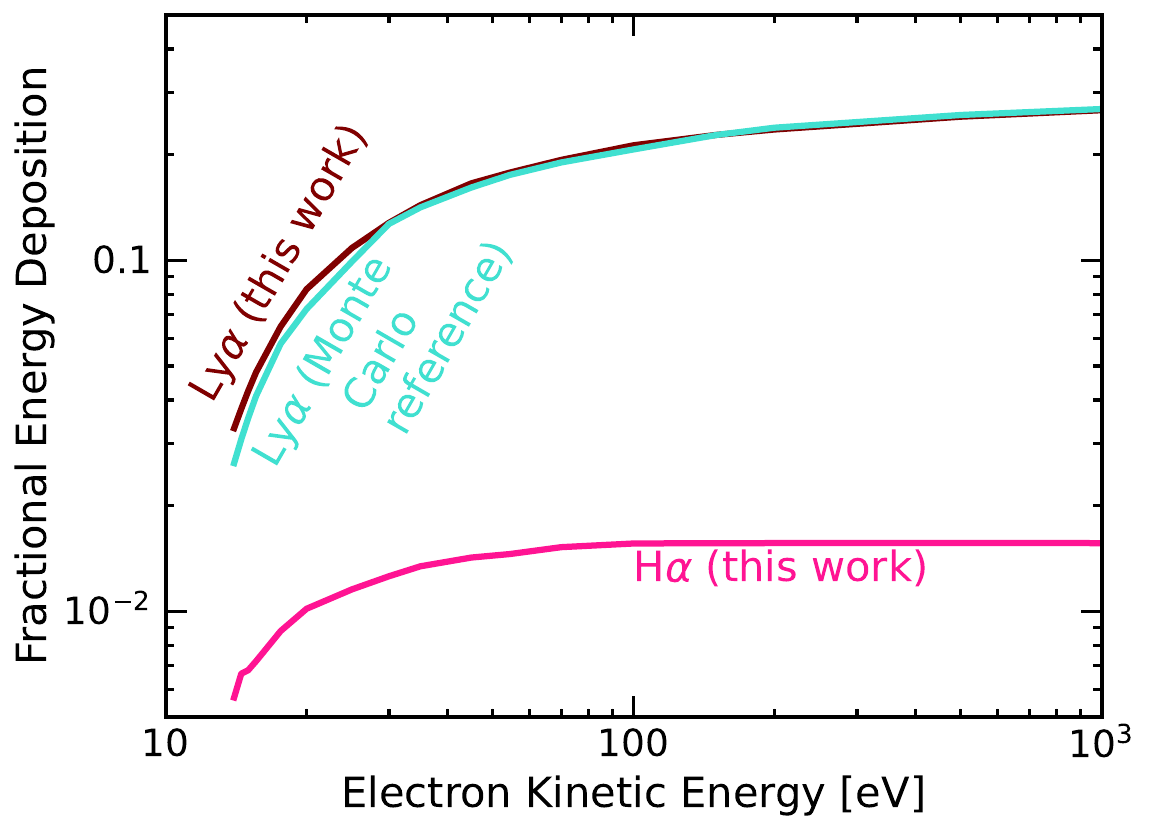}
    \caption{Fraction of electron kinetic energy deposited into $\Ha$- and Ly$\alpha$-producing excitations for a mixed H/He medium with ionization fraction $x_e=0.01$. Agreement of our Ly$\alpha$ result with the Monte Carlo calculation of Ref.~\cite{Furlanetto_2010} validates our treatment of the excitation energy budget.}
    \label{fig:fHa and fLa at xe=0.01}
\end{figure}

Here we provide further details of the atomic-level treatment used for electron-impact excitation and present the resulting fraction of electron energy deposited into $\Ha$-producing excitations. We also validate our method by applying it to Ly$\alpha$ production and comparing with the Monte Carlo results of Ref.~\cite{Furlanetto_2010}.

We use the $n \ell$-resolved electron-impact excitation cross sections $\sigma^{e}_{1s\to n\ell}$ from the convergent close-coupling (CCC) database \cite{CCCDatabase} for all states through $4f$. For excitations to $n\geq5$, we follow Ref.~\cite{Furlanetto_2010} and approximate
\begin{equation}
\sigma_{n\geq5}=1.43\,\sigma_{n=4},
\end{equation}
where $\sigma_{n=4}$ is summed over the angular-momentum substates. We calculate the cascade probability $P_{\Ha}(n\ell)$ explicitly for states through $4f$ and take
\begin{equation}
P_{\Ha}(n\geq5)=P_{\Ha}(n=4),
\end{equation}
where $P_{\Ha}(n=4)$ is the unweighted average over the four $n=4$ sub-levels.

The CCC cross sections used directly in the calculation extend to electron energies below $1~\mathrm{keV}$. At higher energies, we extrapolate the excitation energy-deposition fractions using their Bethe-limit behavior \cite{Bethe1930,RevModPhys.43.297,Furlanetto_2010}.

Figure~\ref{fig:fHa and fLa at xe=0.01} shows the $\Ha$ deposition fraction used in our analysis, for an ionization fraction $x_e=0.01$ appropriate to the Leo~T WNM. As a check, we apply the same semi-analytic procedure to Ly$\alpha$-producing excitations and find agreement with the Monte Carlo calculation of Ref.~\cite{Furlanetto_2010}, which follows the complete cascade of secondary electrons through a partially ionized gas. This validates our treatment of the total excitation energy budget; the $\Ha$ yield additionally depends on the distribution of excited states, which we take from the CCC cross sections.

\section{Axion $\Ha$ Energy Deposition via Photon Decay: Supporting Material}
\label{sec: photon dep}

This section provides supporting details for the photon-decay calculation in the main text. We define the normalized line profile appearing in Eq.~(\ref{eq:photoexcitation-cross-section}), describe the state resolution of the atomic data, and discuss the contribution from direct photo-excitation.

\subsection{Normalized Line Profile}

Atomic photo-excitation resonances are broadened by two effects: the finite lifetimes of the initial and final states, which produce a natural Lorentzian profile, and the velocity dispersion of the H atoms, which produces Doppler broadening. Their convolution gives the Voigt profile used in Eq.~(\ref{eq:photoexcitation-cross-section}). For a transition from state $i$ to state $f$, the Voigt profile is given by  \cite{draine2011physics}
\begin{align}
\phi_{i \to f}^{\mathrm{Voigt}} (\ega) =\frac{1}{\sqrt{2\pi}}
    &\int \frac{dv}{\sigma_v}\, e^{-v^{2}/(2\sigma_v^{2})}\,
    \frac{1}{\pi}\,\frac{\gamma_{fi}/2}{\bigl[\ega -(1-v)(E_f - E_{i})\bigr]^{2}+(\gamma_{fi}/2)^{2}} \, ,
    \label{eq: Voigt line profile}\\
\text{where \,\, } 
    \gamma_{fi} &=\gamma_{if} =\sum_{E_j<E_f} A_{f \to j} + \sum_{E_j<E_i} A_{i \to j} \, . \label{eq: gamma}
\end{align}
Here $\sigma_v$ is the line-of-sight velocity dispersion of the H atoms, and $A_{i\to j}$ is the Einstein coefficient for spontaneous emission from level $i$ to level $j$. The natural width $\gamma_{fi}=\gamma_{if}$, given by the sum of the total radiative decay rates of the two states, is the full width at half maximum of the Lorentzian profile~\cite{draine2011physics}. The profile is normalized such that
\begin{equation}
\int dE_\gamma \,
\phi_{i\to f}^{\rm Voigt}(E_\gamma)=1\,.
\end{equation}
For the Leo~T WNM, we adopt $\sigma_v=7.9~\mathrm{km\,s^{-1}}$, the thermal velocity dispersion at $7552~\mathrm{K}$~\cite{Wadekar:2021qae}; however our results are insensitive to this choice (the DM velocity dispersion is similarly negligible). This is because near line center the gas is optically thick regardless of the Doppler width, while away from line center the profile is dominated by the Lorentzian wings, which are independent of $\sigma_v$.

\subsection{Off-Resonance Absorption and the MUSE Spectral Window}

Near resonance, the Voigt profile provides a good approximation to the full quantum-mechanical Kramers--Heisenberg treatment~\cite{draine2011physics,Lee_2003,Lee_2004}. We model the $3p$ and higher resonances with their Voigt profiles, and retain the contribution at energies between resonances, since such off-resonance photons can still produce $\Ha$ via virtual intermediate states. Away from line center, however, absorption and re-emission proceed as a single scattering event, so by energy conservation the detuning $\Delta E\equiv E_\gamma-E_{1s\to np}$ of the incident photon is carried by the first photon the atom emits (Raman scattering).

This matters because the MUSE limit counts only $\Ha$ within the window used in its derivation: a Gaussian of width $70~{\rm km\,s^{-1}}$, summed over $\pm5$ widths around $\Ha$ at the systemic velocity of Leo~T~\cite{Vaz_2023}, i.e.\ $\pm7.7$~\AA, or $|\Delta E|\leq\Delta E_{\rm win}\simeq2.2$~meV. This window is much broader than the $\sigma_v=7.9~{\rm km\,s^{-1}}$ WNM dispersion that sets our line profiles, so $\Ha$ emitted at line center lies well within it. For the $3p$ resonance, the only $\Ha$-producing decay is $3p\to2s$, so the $\Ha$ photon itself is shifted by $\Delta E$; we therefore count the $3p$ contribution only for $|\Delta E|\leq\Delta E_{\rm win}$, and set it to zero otherwise. For $n\geq4$, $\Ha$ is produced only if the atom first decays to a state with $n\geq3$ (e.g.\ $4p\to3s$, $3d$), emitting an infrared photon. This photon carries the detuning, leaving the atom in a real state that can subsequently emit $\Ha$ at line center. We therefore retain the full Voigt contribution of these resonances.

Between resonances, where the gas is optically thin, we continue to treat Lyman decays as recycled when computing $P_{\Ha}$. In practice, photons scattered back to $1s$ there retain their off-resonance energy and largely escape rather than being reabsorbed, leading to an overestimation by up to a factor of $\sim2$ in $g_{a\gamma\gamma}$. Our limits between resonances should therefore be regarded as approximate; those at line centers, where $\tau\gg1$ and our results are strongest, are unaffected. A more accurate treatment of the off-resonance region would require a full Kramers--Heisenberg analysis, including interference between resonances, which we leave for future work.

\subsection{Mass-Energy Absorption Coefficients Below Ionization Thresholds}

For $E_\gamma<13.6~{\rm eV}$, H cannot be ionized, and we calculate the mass-energy absorption coefficient from bound-bound photo-excitation, $(\mu_{\rm en}/\rho)_{\rm H}=\sum_{n\ell}\sigma^\gamma_{1s\to n\ell}(E_\gamma)/m_{\rm H}$, with cross sections from Eq.~(\ref{eq:photoexcitation-cross-section}). We include states through $n=40$, the highest principal quantum number with data available in Ref.~\cite{NIST_ASD}, corresponding to an excitation energy of $13.59~{\rm eV}$. We use $\ell$-resolved transition data through $n=12$ and $\ell$-averaged data at higher $n$. He does not contribute in this range, as its lowest excited level lies at $19.8~{\rm eV}$.

For $13.6\leq E_\gamma<24.6~{\rm eV}$, H can be ionized but He cannot, and we use the results of Ref.~\cite{1990A&A...237..267B}, which include H photoionization and treat He's mass-energy absorption coefficient as negligible below its ionization threshold. Above $21.2~{\rm eV}$, He photo-excitation adds narrow absorption bands around its resonances. We neglect these, as they span narrow energy intervals, and He re-emits most absorbed photons as resonance photons above $13.6~{\rm eV}$, which then ionize H.

\subsection{Direct Photo-Excitation: Level Treatment and Energy Deposition}

We explicitly calculate the cascade probability $P_{\Ha}(i)$ for photo-excitation to states through $6p$, resolving the $n\ell$ states encountered in each subsequent radiative cascade. Above $n=6$, Ref.~\cite{NIST_ASD} provides only $n$-resolved Einstein $A$ coefficients for the cascade transitions, and we adopt
\begin{equation}
P_{\Ha}(np)=P_{\Ha}(6p)\simeq0.29\,,\quad n\geq7\,,
\end{equation}
which is conservative since $P_{\Ha}(np)$ rises slowly with $n$.

\begin{figure}
    \includegraphics[width=0.5\linewidth]{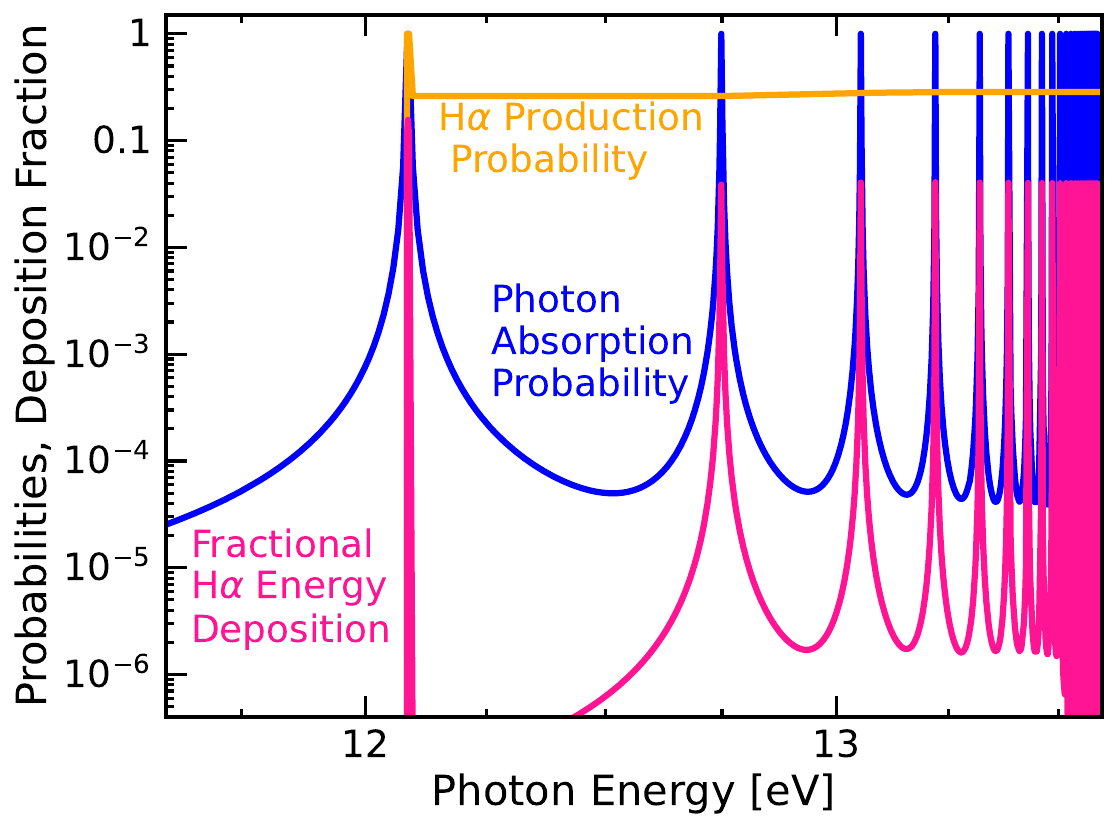}
    \caption{Photon absorption probability $P_{\rm abs}$, $\Ha$ production probability per absorbed photon $P^\gamma_{\Ha}$ (excluding Ly$\beta$-wing absorptions outside the MUSE window), and fraction of photon energy deposited into $\Ha$, $f^{\Ha}_{{\rm exc},\gamma}$, for sub-threshold photons in the Leo~T gas within the MUSE aperture. Only $\Ha$ within the MUSE spectral window is counted; between line centers, the curves are approximate (see text).}
    \label{fig: Direct photo-excitation probabilities}
\end{figure}

Figure~\ref{fig: Direct photo-excitation probabilities} shows the fraction of photon energy deposited into $\Ha$ via direct photo-excitation, Eq.~(\ref{eq: photo-excitation full}). The resonances arise from excitation to discrete $np$ states followed by radiative cascades; from about $10p$ upward, they are closely spaced enough to appear nearly continuous at the resolution shown.

\section{Dark Matter Ionization and Recombination Equilibration}
\label{sec: equilibration}

Here we reproduce for completeness key details of the discussion in Ref.~\cite{Leane:2025ohu} about equilibration between H ionization and recombination, and its impact on axion DM-injected $\Ha$ flux. This is used in the calculation of $f_{{\rm eq},k}$ in Eq.~(\ref{eq:Flux}).

Axion-induced ionization produces free electrons, while H recombination removes them. For axion decay injecting a particle $k\in\{\gamma,e\}$ into the ISM, the free-electron density evolves as
\begin{equation}
  \frac{dn_{e}}{dt} \;=\; {\zeta_{a,k} \,n_{\rm H}}-\alpha_B(T)\,n_{e}^2,
  \label{eq:evolve}
\end{equation}
where $\zeta_{a,k}$ is the axion-induced ionization rate per H atom, $n_{\rm H}$ is the neutral-H density, and $\alpha_B(T)$ is the case-B recombination coefficient. We adopt $\alpha_B\simeq3.16\times10^{-13}~{\rm cm^3 \, s^{-1}}$, evaluated at a conservative Leo T temperature of approximately $7552~{\rm K}$ \cite{1989agna.book.....O,2022MNRAS.510.2797P,RydenPogge2021_IISM}.

For an initially neutral medium, Eq.~\eqref{eq:evolve} approaches equilibrium on the timescale
\begin{equation}
    t_{{\rm eq},k}=\frac{1}{\sqrt{\alpha_B\,\zeta_{a,k} \,n_H}}\,.
\end{equation}
The finite duration over which the local conditions remain approximately steady suppresses the recombination signal relative to its equilibrium value by
\begin{equation}
 f_{{\rm eq},k} \;=\; \tanh^2\!\left(\frac{t_{\rm av}}{t_{{\rm eq},k}}\right)\in(0,1],
  \label{eq:feq}
\end{equation}
where $t_{\rm av}\sim50~{\rm Myr}$ is the effective duration of the quiescent conditions in Leo T \cite{Leane:2025ohu}. In the steady-state limit $t_{\rm av}\!\gg\!t_{{\rm eq},k}$, $f_{{\rm eq},k}\!\to\!1$; for $t_{\rm av}\!\ll\!t_{{\rm eq},k}$ one has $f_{{\rm eq},k}\simeq (t_{\rm av}/t_{{\rm eq},k})^2$. 

As discussed in Section~\ref{sec: Leo T} above, we conservatively neglect any pre-existing ionized component when calculating the equilibration time. Including background free charges would accelerate equilibration and strengthen the resulting limits. For example, an ambient density $n_e^{\rm bg}\sim10^{-3}~{\rm cm}^{-3}$ can increase the ionization contribution by up to approximately an order of magnitude.

In terms of the quantities used in the main text, Eq.~\eqref{eq:feq} becomes
\begin{equation}
f_{{\rm eq},k}
=\tanh^2\!\Big[t_{\rm av}\,\sqrt{\alpha_B\,\zeta_{a,k} \,n_{\rm H}}\;\Big]\,,
\label{eq:feq_def}
\qquad
\text{with} \,\,\,
\zeta_{a,k} = \frac{E_{k}^{\rm dec}\,\Gamma_{a\to kk}}{m_a}\,\frac{\mathcal{D}[A]}{\Omega_A\,L_{\rm depth}} \, \frac{f^{\Ha \rm}_{{\rm ion},k}}{f_{\Ha}E_{\Ha}} \frac{1}{n_{\rm H}}
\end{equation}
where $\mathcal{D}[A]$ is the D-factor defined in Eq.~(\ref{eq:D factor}), replaced by the full-WNM-core D-factor for electrons, $f^{\Ha \rm}_{{\rm ion},k}$ is defined in Eq.~(\ref{eq:Flux}) and calculated in Eqs.~(\ref{eq:fdepe_mono}) and (\ref{eq:fdep_gamma_Ha}), and $\Omega_A$ and $L_{\rm depth}$ correspond to the viewing aperture of MUSE/VLT whose details are in Section \ref{sec: Leo T} above.

\bibliography{main}
\bibliographystyle{apsrev4-2}

\end{document}